\documentclass[11pt,USenglish,letterpaper]{article}
\usepackage[T1]{fontenc}
\usepackage[latin9]{inputenc}
 
\usepackage{microtype}
\usepackage{amsmath}

\usepackage{numprint}
\usepackage{todonotes}
\usepackage{fancyhdr}
\usepackage{placeins}
\usepackage{balance}
\usepackage{hyperref}
\usepackage[margin=0.9in]{geometry}
\usepackage[algo2e,ruled,linesnumbered]{algorithm2e}
\usepackage{babel}
\usepackage{float}
\usepackage{mathtools}
 \usepackage{subcaption}
 \usepackage{amssymb}
 \usepackage{array}
\usepackage{wrapfig}
\usepackage{placeins}
\usepackage{amsmath}
\usepackage{amsthm}
\usepackage{amssymb}

\bibstyle{abbrv}

\usepackage{ifthen}
\newboolean{double_blind}
\setboolean{double_blind}{false}

\newcommand{\ie}{i.\,e.\xspace}

\newcommand{\etal}{et al.\xspace}

\newcount\shortyear\newcount\shorthour\newcount\shortminute
\shorthour=\time\divide\shorthour by 60\shortyear=\shorthour
\multiply\shortyear by 60\shortminute=\time\advance\shortminute by
-\shortyear
\shortyear=\year\advance\shortyear by -1900

\def\zeit{\number\shorthour:\ifnum\shortminute<10 0\number\shortminute
\else\number\shortminute\fi}

\newtheorem{theorem}{Theorem}[section]

\newtheorem{lemma}[theorem]{Lemma}

\date{\today}

\begin{document}

\ifthenelse{\boolean{double_blind}}
{
\author{} 
}
{
\author{Elisabetta Bergamini\footnote{Karlsruhe Institute of Technology (KIT), Germany} \quad Tanya Gonser\textsuperscript{$\dagger$} \quad  H. Meyerhenke\footnote{Humboldt-Universit\"at zu Berlin, Germany -- the previous versions of this paper were
written while the third author was affiliated with Karlsruhe Institute of Technology (KIT), Germany, as well.}
}

\title{Scaling up Group Closeness Maximization\thanks{This work is partially supported by German Research Foundation (DFG) grants ME 3619/3-1 and -2 within the 
Priority Programme 1736 \emph{Algorithms for Big Data.}}}}

\date{}

\maketitle

\begin{abstract}

Closeness is a widely-used centrality measure in social network analysis. For a
node it indicates the reciprocal of the average shortest-path distance to the other nodes of the
network. While the identification of the $k$ nodes with
highest closeness received significant attention, many applications are actually interested in finding a \emph{group}
of nodes that is central as a whole. For this problem, only recently a
greedy algorithm has been proposed [Chen \etal, ADC~2016]. The approximation factor of
$(1-1/e)$ proposed by Chen \etal for this algorithm does not hold, though, as we show in this version of our paper.
Since their implementation of the greedy algorithm was still too slow for large 
networks, Chen \etal also proposed a heuristic without approximation guarantee.

In the present paper we develop new techniques to speed up the greedy algorithm.
Compared to the previous implementation, our approach is orders of magnitude faster and, compared to the heuristic proposed
by Chen \etal, we always find a solution with better quality in a comparable running time in our experiments. 

Our method \textsf{Greedy++} allows us to estimate the group with maximum closeness on networks with up to hundreds of millions of edges in minutes or at most a few hours. The greedy approach by [Chen \etal, ADC~2016] would take several days already on networks with hundreds of thousands of edges.
Our experiments show that the solution found by \textsf{Greedy++} is actually very close to the optimum
(at least on small networks, where we could compute the optimum with an ILP solver).
Over all tested networks, the empirical approximation ratio is never lower than 0.97. 

Finally, as far as we know, we study for the first time the correlation between the top-$k$ nodes with highest individual closeness and an approximation of the most central group in large complex networks. Our results show that the overlap between the two is relatively small, which indicates empirically the need to distinguish clearly between the two problems.

\textbf{Note:} This paper version fixes the issue of relying on the presumed (but incorrect) submodularity of group closeness.
While this has implications on the theoretical assessment of the greedy algorithm, our algorithm variant
and its implementation remain unaffected. The reason is that \textsf{Greedy++} relies (among others) on the supermodularity of farness, which does hold.

\end{abstract}


\section{Introduction}
One of the main tasks in social network analysis is the identification of important nodes. For this reason, numerous centrality measures have been introduced over the years and much work has been put into the efficient computation of centrality scores for individual nodes. Closeness centrality is one of the widely-used measures; it ranks the nodes according to the reciprocal of their average shortest-path distance to the other nodes. Intuitively, a node with high closeness is a node that is close, on average, to the other nodes of the network and can therefore reach them quickly.

In their seminal work, Borgatti and Everett~\cite{doi:10.1080/0022250X.1999.9990219} extended the concept of centrality to \textit{groups} of nodes. For a node $v$ and a group $S$ of other nodes, the distance between $v$ and $S$ is defined as the minimum distance between $v$ and the elements of $S$. Then, a group of nodes has high closeness when its average distance to the other nodes is small.
Finding central groups of nodes is an important task for many applications. For example, in social networks, retailers might want to select a group of nodes as promoters of their product, in order to maximize the spread among users~\cite{DBLP:conf/kdd/KempeKT03}. In this context, picking the $k$ most central nodes might lead to a large overlap in the set of influenced nodes, whereas there might be $k$ nodes that are not among the most central when considered individually, but that influence different areas of the graph. 

Related to finding the group with highest closeness is $p$-median, a fundamental facility location problem in operations research~\cite{DBLP:journals/ior/Hakimi65}. While the standard GCM formulation applies only to graphs without vertex weights, $p$-median (also) applies to geometric inputs and weighted objects (to name only few of the possible generalizations~\cite{zvi}).
For $p$-median, several (meta)heuristics and approximation algorithms have been proposed over the years (see~\cite{DBLP:journals/networks/Reese06} for an annotated bibliograohy). Yet, these methods are mostly applicable to relatively small networks only. In~\cite{DBLP:conf/evoW/RebreyendLE15}, the authors compare state-of-the-art methods on a street network of Sweden ($\approx 190$K nodes) and show that existing methods either fail due to their memory requirements (>32 GB) or take more than 14 hours to find an approximation. Other recent methods have been shown to scale to inputs with up to \numprint{90000} points/nodes~\cite{DBLP:journals/cor/AvellaBSV12, DBLP:journals/jgo/IrawanS15}.

Specifically for GCM, a greedy algorithm has been proposed recently by Chen \etal~\cite{DBLP:conf/adc/ChenWW16}.
Its presumed approximation ratio of $(1-1/e)$ does not hold, though -- as we show in this version of the paper.
(In previous versions of this paper, we based our approximation guarantee assumptions on their incorrect proof sketch -- an issue we have fixed now.)
The greedy algorithm's implementation in Chen \etal does not scale easily to graphs with more than about $10^4$ vertices, since it requires to compute pairwise distances. Thus, its authors proposed in the same paper also a more scalable heuristic (without any guarantees on the solution quality).


\paragraph{Outline and contribution.} 
We present techniques that can reduce considerably the memory and the number of operations required by the greedy algorithm presented in~\cite{DBLP:conf/adc/ChenWW16}. 
First, instead of computing and storing all pairwise distances, we use the algorithm presented in~\cite{DBLP:conf/alenex/BergaminiBCMM16} to find the node with maximum closeness (Section~\ref{sec:greedy}). Then, we reduce the subsequent computations using pruned SSSPs (Section~\ref{sec:pruned}) and exploiting the supermodularity of the objective function farness -- the reciprocal of closeness (Section~\ref{sec:submodularity}). We also propose an approach based on bit vectors (Section~\ref{sec:bitwise}) which is faster than pruned SSSPs but requires more memory in general.
In our experiments in Section~\ref{sec:experiments}, we compare our algorithm (\textsf{Greedy++}) with the greedy approach presented in~\cite{DBLP:conf/adc/ChenWW16} and show that \textsf{Greedy++} is orders of magnitude faster. Also, we compare \textsf{Greedy++} with the heuristic proposed in~\cite{DBLP:conf/adc/ChenWW16} and show that \textsf{Greedy++} is often faster (or has a comparable running time) and that it always finds a better solution in all our experiments.
We also provide an Integer Linear Programming (ILP) formulation of the GCM problem in Section~\ref{sec:ilp} and compare the quality of our solution with the optimum. Our results show that the solution found by \textsf{Greedy++} is close to the optimum: the empirical approximation ratio is never lower than 0.97.
Finally, we study the overlap between the group with maximum closeness and the $k$ nodes with highest closeness and highest degree in real-world networks, showing that in most cases this is relatively small (between 30\% and 60\% of the group size). This confirms the intuition that a central group of nodes is not necessarily composed of nodes that are individually central.


\section{Preliminaries}
We model a network as a graph $G = (V, E)$ with $|V| =: n$ nodes and $|E| =: m$ edges. Unless stated explicitly, we assume the graph to be connected (or, if directed, strongly connected) and unweighted.
Let $d(u, v)$ represent the shortest-path distance between node $u$ and node $v$. We define the distance between $u \in V$ and a set $S \subseteq V$ of nodes as 
$d(u, S) := \min_{s \in S} d(u, s)\ .$
Then, the closeness centrality of node $u$ is defined as
$c(u) := \frac{n-1}{\sum_{v \neq u} d(u, v)}\ .$
Similarly, one could define the closeness of a set $S$ as 
$c(S) := \frac{n-|S|}{\sum_{v \notin S} d(S, v)}\ .$
In line with previous work~\cite{DBLP:conf/adc/ChenWW16}, we omit the normalization in the numerator with respect to the group size.
Thus, we define the group closeness of $S$ as $c(S) := \frac{n}{\sum_{v \notin S} d(S, v)}\ .$
Also, let \emph{group farness} $f$ be the reciprocal of group closeness: $f(S) := 1/f(S)$.

The Group Closeness Maximization (GCM) problem is defined as finding a set $S^{\star} \subseteq V$ of a given size $k$, with maximum group closeness: 
$S^{\star} = {\arg \max}_{S \subseteq V} \{c(S) : |S| = k  \}$.
Note that an adaptation of our approximation algorithms to the case of $\vert S \vert \leq k$ is rather straightforward.
In the paper we use SSSP to denote a single-source shortest path computation, \ie, breadth-first search (BFS) for unweighted graphs. We use APSP to denote an all-pairs shortest path distance computation.

\section{Related work}
Computing closeness centrality requires the distances between all pairs of nodes. For this problem one typically solves a SSSP from each node or uses techniques based on fast matrix multiplication. In both cases the time required is at least quadratic in the number of nodes.
For this reason, several approximation algorithms for closeness centrality have been proposed~\cite{DBLP:journals/jgaa/EppsteinW04,DBLP:journals/ijbc/BrandesP07,DBLP:conf/cosn/CohenDPW14,DBLP:conf/approx/ChechikCK15}. The basic idea is to sample a set of nodes (pivots), compute the distance between the pivots and the other nodes and then estimate the closeness scores of all nodes using the computed distances. 
Although these algorithms can often approximate the scores well, they may fail at preserving the ranking of nodes, in particular for those with similar closeness values. In~\cite{DBLP:journals/corr/BergaminiBCMM17}, it has been shown that the algorithm by Chechik \etal~\cite{DBLP:conf/approx/ChechikCK15} would require $n^2$ SSSP computations to guarantee an exact ranking in complex networks, which is clearly impractical.
For this reason, recently \textit{exact} algorithms for finding the $k$ nodes with maximum closeness have been proposed~\cite{DBLP:conf/alenex/BergaminiBCMM16,DBLP:journals/corr/BergaminiBCMM17,DBLP:journals/corr/BorassiCM15,DBLP:conf/icde/OlsenLH14}. 
The authors of~\cite{DBLP:journals/corr/BorassiCM15} propose an algorithm with a worst-case complexity of $\Theta(nm)$; in practice, however, it appears to be very scalable. Subsequently, the algorithm presented in~\cite{DBLP:journals/corr/BorassiCM15} has been further improved in~\cite{DBLP:conf/alenex/BergaminiBCMM16} and extended in~\cite{DBLP:journals/corr/BergaminiBCMM17}. Since we use this algorithm to solve a subtask of our greedy approach for group closeness maximization, we describe it in Section~\ref{sec:bfscut}.

GCM has been recently considered in~\cite{DBLP:conf/adc/ChenWW16}; 
the authors sketch a hardness proof for GCM with a reduction from an NP-hard clustering problem. Also, they propose a greedy algorithm which they thought
to be an approximation algorithm with factor $(1-1/e)$. The proof sketch for this approximation guarantee is incorrect, however -- due to an invalid
assumption on the relation of submodularity and supermodularity (see Section~\ref{sec:greedy}). 
Since the greedy algorithm is still expensive (its complexity is $\Theta(k n^2)$ plus the cost of an APSP, for a group of size $k$), the authors propose an alternative heuristic based on sampling. In particular, they first propose a baseline heuristic (\textsf{BSA}), which basically samples a set of nodes and then selects iteratively the node that minimizes the distance of the current solution to the samples. Then, they show that the running time of \textsf{BSA} can be improved by dividing the set of samples in partitions (and they call this second heuristic Order-based Sampling Algorithm, \textsf{OSA}).
There is also no guarantee known on the solution quality of the latter two heuristics.
Since the algorithm proposed in this paper builds on the greedy algorithm of~\cite{DBLP:conf/adc/ChenWW16}, we describe it in more detail in Section~\ref{sec:greedy}.

In~\cite{DBLP:conf/www/ZhaoLTG14}, an algorithm for computing and maximizing group closeness on disk-resident graphs has been proposed. The basic idea is to estimate the closeness of a group using the nodes at distance at most $H$ from the group (where $H$ can be any integer value greater than 0). Although they show that their approach can scale quite well for small values of $H$, there is no guarantee on how close their estimation is to the real centrality of the group.

The problem of finding a central group of nodes has also been considered for betweenness centrality, for which sampling-based approximation algorithms have been proposed~\cite{DBLP:conf/kdd/MahmoodyTU16,DBLP:conf/kdd/Yoshida14}.

\subsection{Top-$k$ closeness algorithm}
\label{sec:bfscut}
The basic idea of the top-$k$ closeness algorithm for complex networks proposed in~\cite{DBLP:conf/alenex/BergaminiBCMM16} can be summarized as follows: Let us assume we want to find the $k$ nodes with highest closeness centrality. Also, assume we have an upper bound $\tilde{c}(v)$ on the closeness of a node $v$. Then, if $k$ nodes exist such that their exact closeness is higher than the upper bound $\tilde{c}(v)$, we know that $v$ is not one of the $k$ nodes with highest closeness and we do not need to compute its exact closeness $c(v)$. 
The algorithm is summarized in Algorithm~\ref{alg:topk} in Appendix~\ref{app:pseudocodes}. In each iteration, $x_k$ contains the $k$-th highest closeness value found so far. 
Function $\mathtt{BFScut}(v, x_k)$ in Line~\ref{line:bfscut} computes iteratively an upper bound on the closeness of $v$ in the following way: A BFS rooted in $v$ is initiated. After all nodes up to a certain distance $d$ from $v$ have been visited, we know that all remaining nodes are \textit{at least} at distance $d+1$. If we assume that all the unvisited nodes are \textit{exactly} at distance $d+1$, this gives us an upper bound $\tilde{c}_d(v)$ on the closeness of $v$ for each possible distance value $d$. Therefore, at each step of the BFS rooted in $v$, we can compare $\tilde{c}_d(v)$ with $x_k$. If $x_k \geq \tilde{c}_d(v)$, then we can interrupt the BFS and return 0, meaning that $v$ is not one of the top-$k$ nodes. Otherwise, a whole BFS is computed for $v$ (and \texttt{BFScut} returns the exact closeness of $v$).
Function $\mathtt{Kth}(c)$ in Line~\ref{line:kth} returns the $k$-th largest element of $c$ and $\mathtt{TopK}(c)$ in Line~\ref{line:topk} returns the $k$ largest elements of $c$.
%
%
In~\cite{DBLP:conf/alenex/BergaminiBCMM16}, some improvements on Algorithm~\ref{alg:topk} are proposed. The idea is to compute upper bounds on the closeness of each node in a pre-processing phase and then process the nodes according to these bounds instead of their degree. For more details, we refer the reader to~\cite{DBLP:conf/alenex/BergaminiBCMM16}.

\subsection{Greedy algorithm}
\label{sec:greedy}
Chen \etal~\cite{DBLP:conf/adc/ChenWW16} proposed a greedy algorithm (\textsf{Greedy}) for group closeness.
We recall that the objective is to find a set $S^{\star}$ such that $S^{\star} = {\arg \max}_{S \subseteq V} \{c(S) : |S| = k  \}. $
\textsf{Greedy} runs $k$ iterations, after which it returns a set $S$. Within each iteration, \textsf{Greedy} adds to the set $S$ the node $u$ with the largest \emph{marginal gain} $c(S \cup \{u\}) - c(S)$.
Since $c(S)$ does not depend on $u$, this corresponds to finding the node $u$ with maximal $c(S \cup \{u\})$.
Actually, we express and implement the algorithm in terms of farness minimization; this equivalent view then means
to search for the node $u$ with minimal $f(S \cup \{u\})$ in each iteration.
Expressing the algorithm in terms of farness has the advantage that we can exploit
the property of supermodularity in Section~\ref{sec:submodularity}.

A set function $h$ is \emph{supermodular} if $h(S \cup \{u\}) - h(S) \leq h(T \cup \{u\}) - h(T)$, for any two sets $S \subseteq T$ from some common ground set $V$ and any $u \in V \setminus T$. For \emph{submodularity} the inequality has to be reversed.
Group farness $f$ of a set $S$ is a supermodular function. This was already observed by 
Chen \etal~\cite{DBLP:conf/adc/ChenWW16}; we provide a more detailed proof below.
\begin{lemma}
Group farness $f(S)$ for $S \subseteq V$ is a supermodular set function. 
\end{lemma}
\allowdisplaybreaks
\begin{proof}
Let $S \subseteq T \subseteq V$ and $v \in V \setminus T$. We use the notation
$V^-_{T,v} := \{w \in V \setminus T: d(T \cup \{v\}, w) < d(T, w)\}$ for the set of nodes $w$ closer 
to $v$ than to $T$. 
Then:
\begin{align*}
    f(T \cup \{v\}, w) - f(T) & = \frac{1}{|V|} \left(\sum_{w \in V \setminus (T \cup \{v\})} d(T \cup \{v\}, w) - \sum_{w \in V \setminus T} d(T, w) \right)\\
& = \frac{1}{|V|} \left( \sum_{w \in V \setminus T} d(T \cup \{v\}, w) - d(T, w) \right) \\
& = \frac{1}{|V|} \left( \sum_{w \in V^-_{T,v}} d(T \cup \{v\}, w) - d(T, w) \right)  \\
& = \frac{1}{|V|} \left( \sum_{w \in V^-_{T,v}} d(S \cup \{v\}, w) - d(T, w) \right) \\
& \geq \frac{1}{|V|} \left( \sum_{w \in V^-_{T,v}} d(S \cup \{v\}, w) - d(S, w) \right) \\
& \geq f(S \cup \{v\}) - f(S).
\end{align*}
The first equation follows from the definition of group farness, the second equation uses $d(T \cup \{v\}, v) = 0$.
Then, the third equation uses the definition of $V^-_{T,v}$; the respective contribution of all other summands vanishes.
The fourth equation follows from $d(T \cup \{v\}, w) = d(v, w) = d(S \cup \{v\}, w)$ for all
$w \in V^-_{T,v}$ by definition. Finally, we use $d(S, w) \geq d(T, w)$ and that additional 
non-positive summands for nodes $w$ from a larger set cannot make the total sum larger.
\end{proof}

In their submodularity proof for group closeness, Chen \etal made the implicit (but wrong) assumption that the supermodularity of $f$ implies submodularity
of group closeness $c$. Since they also showed that closeness is monotonic,
this led them (and in the following also us in previous versions of this paper) to believe 
that \textsf{Greedy} would provide a $(1 - 1/e)$-approximation for the GCM problem.

However, submodularity does not hold for closeness. Consider $K_5$, the complete graph with 5 nodes,
numbered from $0$ to $4$. Let $S = \{0, 1\}$, $T = \{0, 1, 2\}$ and $v = 3$. Then:
\begin{equation*}
c(S \cup \{v\}) - c(S) = \frac{5}{2} - \frac{5}{3} = \frac{5}{6} \\
< c(T \cup \{v\}) - c(T) = \frac{5}{1} - \frac{5}{2} = \frac{5}{2}.  
\end{equation*}

Still, a greedy method is a viable approach for the problem at hand. 
Greedy supermodular minimization by starting with $S = V$ and \emph{removing} the worst node 
in each iteration is considered by Il'ev~\cite{ILEV2001131}; he provides approximation results in terms of steepness.
Yet, due to its high number of iterations ($n-k$) for smaller $k$ (small values are more relevant
in practice), we keep the original \textsf{Greedy}
framework that starts with $S = \emptyset$ and \emph{adds} the node with optimal marginal gain.
Algorithm~\ref{alg:greedy} in Appendix~\ref{app:pseudocodes} shows the pseudocode of this latter 
\textsf{Greedy} variant. In Line~\ref{line:distances} the pairwise distances are computed and stored in the $n \times n$ matrices $d$ and $M$. In each iteration, $d$ always contains the pairwise distances, whereas $M$ contains, for each node pair $(u, w)$, the distance $d(S \cup \{u\}, w)$.
Initially $d = M$, since $S = \emptyset$. Then, every time a node $s$ is added to $S$, $M$ is updated in Line~\ref{line:updateM}. $Score$ contains $c(S \cup \{u\})$ for each node $u$, which is computed in Line~\ref{line:updateScore} by summing over $M(u, w)$, $\forall w \in V$.

Since it needs to store two $n \times n$ matrices, the memory requirement of \textsf{Greedy} is $\Theta(n^2)$. The running time is $\Theta(n(m + n \log n))$ for the initial APSP computation (when running a SSSP from each node in a weighted graph) and then $\Theta(k n^2)$ for the remaining part.

\section{A scalable greedy algorithm}
\label{sec:algo}
First of all, we notice that we can reduce the memory requirement of \textsf{Greedy} by not storing the matrices $d$ and $S$. 
In fact, to find the first element $s_0$ of $S$ (\ie the node with maximum closeness) we can simply use the \textsf{TopKCloseness} algorithm described in Section~\ref{sec:bfscut}. Then, we can use a vector $d_S$ containing, for each node $v$, the distance between $S$ and $v$ (i.e. $d_S[v] := d(S, v)$). 
Since initially $S$ is composed of only one element $s_0$, $d_S$ simply contains the distances between $s_0$ and the other nodes, which can be computed with a SSSP rooted in $s_0$.  
Then, Lines~\ref{line:for1}-\ref{line:updateM} can be replaced with a SSSP rooted in $u$ where we sum, over each node $w$ visited in the SSSP, the minimum between $d_S(w)$ and $d(u, w)$. This sum is exactly the same as $\sum_{w \in V\setminus S} M[u, w]$ and can therefore be used in Line~\ref{line:updateScore} to update $Score[u]$.
The memory-efficient version of \textsf{Greedy} is described in Algorithm~\ref{alg:mem-greedy} in Appendix~\ref{app:pseudocodes}. In the pseudocode we report explicitly every time we need to run a SSSP. In Line~\ref{line:sssp0} and Line~\ref{line:sssp2}, the SSSP is needed to compute $d_S$, whereas in Line~\ref{line:sssp} we need it to compute $Score[u]$.

 Since we have to re-run a SSSP for each node $u$ and for each element of $S$ other than $s_0$, the running time complexity of the while loop of Algorithm~\ref{alg:mem-greedy} is $O(kn (m + n \log n))$ (for weighted graphs). The worst-case complexity of finding $s_0$ with \texttt{TopCloseness} is the same as that of an APSP (\ie $n (m +n \log n )$), although in practice it was shown to be basically linear in the size of the graph~\cite{DBLP:journals/corr/BergaminiBCMM17}. For unweighted graphs, the complexity of Algorithm~\ref{alg:mem-greedy} is $O(knm)$, since we can use BFS instead of Dijkstra to compute the SSSPs. 
  Although the memory requirement is now only $\Theta(n)$ (in addition to the memory required to store the graph), the time complexity is too high to target large networks. For this reason, in the following we propose improvements that, as we will see in Section~\ref{sec:experiments}, increase the (sequential) scalability of \textsf{Greedy} considerably.
\subsection{Pruned SSSP}
\label{sec:pruned}
\begin{figure}[t]
\centering
\includegraphics{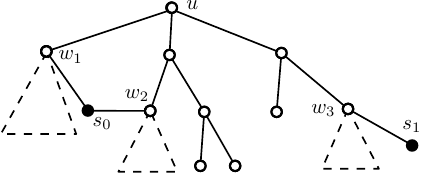}
\caption{Pruned SSSP. If a node $w$ is such that $d_S[w] \leq d(u, w)$, the same holds for the whole SSSP subtree rooted in $w$. In the figure, black nodes represent elements of $S$.}
\vskip -6pt
\label{fig:pruned-sssp}
\end{figure}

In Line~\ref{line:sssp} of Algorithm~\ref{alg:mem-greedy}, we need to run a SSSP rooted in $u$ to recompute $Score[u]$. However, the only nodes $w$ for which we need to compute $d(u,w)$ are those for which $d(u, w) < d_S[w]$, \ie the ones that are closer to $u$ than to $S$. Indeed, for all the other nodes, the distance from $u$ does not contribute to the sum in Line~\ref{line:sumScore} and therefore to $Score[u]$.
Thus, if we know that $d(u,w)$ is larger than or equal to $d_S[w]$, we do not need to visit $w$ in the SSSP. 
It is not hard to see that, if $d(u,w) \geq d_S[w]$, then the same holds for \textit{all the nodes} in the SSSP subtree rooted in $w$. In fact, let $t$ be a node in the SSSP subtree of $w$, \ie $d(u, t) = d(u, w) + d(w, t)$.  There is a path between a node in $S$ and $t$ going through $w$ of length $d_S[w] + d(w, t)$. Therefore $d_S[t] \leq d_S[w] + d(w, t) \leq d(u, t)$. Figure~\ref{fig:pruned-sssp} illustrates this concept.
This allows us to \textit{prune} the SSSP when we find a node whose distance from $u$ is not smaller than its distance from $S$. When we visit a new node $w$, we compare $d(u, w)$ with $d_S[w]$. If the first is not strictly smaller than the second, we do not enqueue its neighbors into the SSSP (priority) queue.
Notice that, since only nodes $u$ for which $d_S[u] \leq d(s, u)$ are pruned, the value of $d_S[u]$ in Line~\ref{line:compdS} is not affected, for any $u \in V$. This means that the solution returned by the improved algorithm is exactly the same as the solution returned by Algorithm~\ref{alg:mem-greedy}.

\subsection{Improvement by exploiting supermodularity}
\label{sec:submodularity}
We can use the supermodularity of farness to reduce the number of evaluations of $Score$ (and SSSP computations) in Lines~\ref{line:sssp}-\ref{line:actScore}, similar to the ``accelerated greedy algorithm'' by Minoux~\cite{10.1007/BFb0006528}. 
Let us name $S_i$ the set $S$ computed by Algorithm~\ref{alg:mem-greedy} in the $i$-th iteration of the while loop ($S_i$ is the set composed of $i$ elements).
Since $S_i \subseteq S_{i+1}$, because of supermodularity $f(S_i \cup \{ u \}) - f(S_i ) \leq f(S_{i+1} \cup \{ u \}) - f(S_{i+1} )$. The difference $f(S_i \cup \{ u \}) - f(S_i )$ is then the \textit{marginal gain} $\Delta(u, S_i)$ ($\Delta_i(u)$, in short) of $u$ with respect to $S_i$.
(Note that the optimal node $u$ to add would be the same if we used the formulation in terms of closeness since
the $f(S_i)$ does not depend on $u$.)
In other words, we can say that in each iteration of the while loop in Algorithm~\ref{alg:mem-greedy}, the marginal gain of each node can only increase. 

Now, let us assume that there is a node $s$ whose marginal gain $\Delta_i(s)$ with respect to $S_i$ is smaller (or equal) than the marginal gain $\Delta_{i-1}(u)$ of a node $u$ in the previous iteration, \ie $\Delta_i(s) \leq \Delta_{i-1}(u)$.
This means that the marginal gain of $u$ at iteration $i$ cannot be smaller than $\Delta_i(s)$ (since $\Delta_i(u) \geq \Delta_{i-1}(u) \geq \Delta_{i}(s)$). This allows us to skip the computation of the score of $u$ in Lines~\ref{line:sssp} - \ref{line:actScore}. All we need to do is keep track of the marginal gain of each node in the previous iteration and compare it with the smallest marginal gain found in the current iteration.
Note that this improvement is compatible with the pruned SSSP improvement proposed in the previous section. For the nodes that cannot be skipped because of what was described in this section, we compute their score with a pruned SSSP.
We name our version of Algorithm~\ref{alg:mem-greedy} using pruned SSSPs and the supermodularity improvement \textsf{Greedy++}. 
As explained in Section~\ref{sec:pruned}, using pruned SSSPs does not affect the solution found by the algorithm. The improvement described in this section can only return a different solution in case there are nodes with the same marginal gain. Indeed, if there are two nodes $u$ and $v$ with the same marginal gain $\Delta^{\star}$ and such that $\Delta^{\star} \leq \Delta(w) \ \forall w \in V$, whether we choose $u$ or $v$ depends on which comes first in the ordering of the nodes.


\newcommand{\vect}[1]{\mathbf{#1}}

\subsection{Bit-parallel group closeness}
\label{sec:bitwise}
To further speed up \textsf{Greedy++}, we propose an optimization for unweighted graphs exploiting bit-level parallelism. Bit-parallel methods try to exploit the fact that computers can perform bitwise operations on
a word at once. Let $\vect{b}^i_u$ be a bit vector with the $j$-th bit set to $1$ if $d(u, j) \leq i$ and set to $0$ otherwise. It is easy to see that $\vect{b}^i_u = \bigoplus_{v \in N(u)} \vect{b}^{i-1}_v$, for $i \geq 1$, where $\oplus$ represents a bitwise-OR operation and $N(u)$ are the neighbors of $u$. Then, if we indicate the number of ones in a bit vector $\vect{b}$ as $|\vect{b}|$, the closeness $c(u)$ of $u$ can be expressed as $\sum_{i=1}^{\mathsf{diam}} i (|\vect{b}^i_u| - |\vect{b}^{i-1}_u|)$, where $\mathsf{diam}$ is the diameter of $G$. A simple algorithm for computing the closeness of all nodes could therefore work as follows: Initialize $\vect{b}^0_u$ as a bit vector with a $1$ in position $u$ and $0$ everywhere else, for each $u \in V$. Then, for $i = 1, \ldots, \mathsf{diam}$, compute  $\vect{b}^i_u$ as  $ \bigoplus_{v \in N(u)} \vect{b}^{i-1}_v$. Although the complexity of this algorithm ($O(\mathsf{diam} \cdot n m)$) is higher than that of running a BFS from each node ($O(nm)$), it can be worthwhile: \textsf{diam} is usually very small in complex networks and bitwise operations are very fast (see for example~\cite{DBLP:conf/ipps/SariyuceSKC14}). 

We can use bitwise operations also to compute \emph{group} closeness. Similarly to $\vect{b}^i_u$, we can define $\vect{b}^{i}_S$ of a set $S$ as a bit vector where the $j$-th bit set to $1$ iff $d(S, j) \leq i$. Then, using $\odot$ to indicate bitwise-AND, and $\lnot$ for bitwise-NOT, we can prove the following:
\begin{theorem}
\label{theo:bitwise}
The node $u^{\star}$ with the highest marginal gain with respect to set $S$ and group closeness is 
\[
u^{\star} = \arg\max_{u \in V \setminus S} \sum_{i = 0} ^ {\mathsf{maxD}(u)} |\vect{b}^i_u \odot \lnot \vect{b}^i_S|,
\]
where $\mathsf{maxD}(u) := \max \{i \geq 0 : |\vect{b}^i_u \odot \lnot \vect{b}^i_S| > 0\} $.
\end{theorem}
\begin{proof}
Recall that it is equivalent to search either for the largest marginal closeness gain or for the smallest
marginal farness gain. This proof relies on closeness. Hence, the marginal closeness gain $\Delta(u, S)$ of node $u$ with respect to set $S$ is $c(S \cup \{u \}) - c(S)$.

Clearly, $\Delta(u, S) > \Delta(v, S) \iff \sum_{w \in V} (d(S, w) - d(S \cup \{u\}, w)) >  \sum_{w \in V} (d(S, w) - d(S \cup \{v\}, w))$, for any two nodes $u$ and $v$. Now, naming $V(u)$ the set of nodes $w$ such that $d(S, w) > d(u, w)$, we can write $\sum_{w \in V} (d(S, w) - d(S \cup \{u\}, w))$ as $\sum_{w \in V(u)} (d(S, w) - d(u, w))$. Thus, the node $u^{\star}$ with maximum marginal gain is $\arg \max_{u \in V \setminus S} \sum_{w \in V(u)} (d(S, w) - d(u, w))$.

For $i \geq 0$, $|\vect{b}^i_u \odot \lnot \vect{b}^i_S|$ is the number of nodes $w$ such that $d(u, w) \leq i$ (as they are in $\vect{b}^i_u$) and $d(S, w) > i$ (as they are not in $\vect{b}^i_S$). Let $w$ be any node in $V(u)$. For each $i$ such that $d(u, w) \leq i < d(S, w)$, the bit corresponding to $w$ in $\vect{b}^i_u \odot \lnot \vect{b}^i_S$ is set to 1. This means that, for each $w \in V(u)$,  $\sum_{i = 0} ^ {\mathsf{maxD}(u)} |\vect{b}^i_u \odot \lnot \vect{b}^i_S|$ adds one to the sum $(d(S, w) - d(u, w))$ times. This means that $\sum_{i = 0} ^ {\mathsf{maxD}(u)} |\vect{b}^i_u \odot \lnot \vect{b}^i_S| = \sum_{w \in V(u)} (d(S, w) - d(u, w))$, which proves the theorem.
\end{proof}

Theorem~\ref{theo:bitwise} gives us a simple algorithm for finding the node with maximum marginal closeness gain: First, we compute $\vect{b}^i_S$, for $i \leq \textsf{diam}$. Then, for each distance $i$ starting from 1 and for each node $u$, we compute $\vect{b}^i_u$ as  $ \bigoplus_{v \in N(u)} \vect{b}^{i-1}_v$. Notice that, if $|\vect{b}^i_u \odot \lnot \vect{b}^i_S| = 0$ for some value of $i$, this will also be true for any $j > i$, so the search from $u$ can be interrupted at distance $i$. This is in some sense equivalent to the pruned SSSP described in Section~\ref{sec:pruned}, but using bit vectors. Also, notice that the algorithm can be combined with the supermodularity improvement in Section~\ref{sec:submodularity}. 

Although using bit vectors can speed up the algorithm (up to a factor 4 in our experiments in Section~\ref{sec:exp-bitwise}), a major limitation of this approach is its memory requirement: for each node, we need to store a bit vector of length $n$, leading to a total of $\Theta(n^2)$ memory. This yields a tradeoff between memory and speed.
It is certainly possible to fathom this tradeoff in more detail. One could specify a certain memory limit not
to be exceeded. Whenever the algorithm needs more memory than the limit, intermediate results need to be
aggregated, so that bit vectors can be reused. This would remove some of its time advantage, however. We regard a detailed investigation of this approach and tradeoff as beyond the scope of this paper and leave it for future work.

\section{ILP formulation of group closeness}
\label{sec:ilp}
To evaluate the quality of the solution found by \textsf{Greedy++}, we want to know how far it is from the optimum. Computing the closeness centrality of all possible subsets of size $k$ would clearly be prohibitive even for tiny networks.
Hence, we formulate GCM as an ILP problem. This will be used in the experiments in Section~\ref{experiments:ilp}.

For each node $v_j \in V$, we define a binary variable $y_j$, which is 1 if node $v_j$ is part of the group with maximum closeness $S^{\star}$, and is equal to 0 otherwise. 
We say a node $v_i$ is \textit{assigned} to a node $v_j \in S^{\star}$ if $d(v_i, S^{\star}) = d(v_i, v_j)$. If there are multiple nodes $v_j \in S^{\star}$ that satisfy this property, $v_i$ can be arbitrarily assigned to one of them.
Thus, we also define a variable $x_{ij}$ that, for each node pair $(v_i, v_j)$ is equal to 1 if $v_j \in S^{\star}$ and $v_i$ is assigned to $v_j$, and 0 otherwise. We can rewrite our problem in the following form:
 \begin{equation}
 \label{eq:lp}
 \max \frac{n}{\sum_{i = 1}^n \sum_{j = 1}^n d(v_i, v_j) x_{ij}}
 \end{equation}
s.t.:
$(i)\sum_{j = 1}^{n} x_{ij} =1, \ \forall i \in \{1,...,n\}$; 
$\ (ii) \sum_{j = 1}^{n} y_{j} =k$; 
$\ (iii)\  x_{ij} \leq y_j,\ \forall  i, j \in \{1,...,n\}.$
\vspace{0.5ex}

Condition $(i)$ indicates that each node in $v_i \in V$ is assigned to exactly one node in $v_j \in S^{\star}$, $(ii)$ indicates that $|S^{\star}| = k$ and $(iii)$ indicates that nodes $v_i$ are only assigned to nodes $v_j$ that are in $S^{\star}$, \ie nodes for which $y_j = 1$. Since the numerator in Eq.~(\ref{eq:lp}) is constant, we can rewrite Eq.~(\ref{eq:lp}) as:
 \begin{equation}
 \label{eq:lp2}
 \min \sum_{i = 1}^n \sum_{j = 1}^n d(v_i, v_j) x_{ij},
 \end{equation}
which gives us an ILP formulation. Values reported in this paper are based on Eq.~(\ref{eq:lp2}).

\section{Experiments}
\label{sec:experiments}
In the following, we present experimental results concerning several aspects of our new algorithm \textsf{Greedy++}. Apart from Section~\ref{sec:exp-bitwise} (where we compare the two versions), we always refer to the version using pruned SSSPs described in Section~\ref{sec:pruned} and not to the one using bit vectors described in Section~\ref{sec:bitwise}.
In Section~\ref{experiments:ilp} we study the accuracy of \textsf{Greedy++} in comparison with the optimum. In Section~\ref{sec:speedup}, we show the speedup of \textsf{Greedy++} on the greedy algorithm proposed in~\cite{DBLP:conf/adc/ChenWW16} (which we call \textsf{Greedy}). 
Then, in Section~\ref{sec:osa}, we compare \textsf{Greedy++} with \textsf{OSA}, the heuristic based on sampling proposed in~\cite{DBLP:conf/adc/ChenWW16} (we did not implement the other heuristic \textsf{BSA}, since the authors of~\cite{DBLP:conf/adc/ChenWW16} show that \textsf{OSA} always finds a solution with a similar accuracy as \textsf{BSA} in a shorter running time).
In Section~\ref{sub:running-time}, we study the running time of \textsf{Greedy++} on additional larger networks, both for a sequential and a parallel implementation
(the other algorithms are either too slow or would require too much memory for these networks).
 Finally, in Section~\ref{sec:group-vs-top}, we study the correlation between the group with maximum closeness and the top-$k$ nodes with highest closeness in real-world networks.

All algorithms are implemented and available in C++ as part of the open-source network analysis tool NetworKit~\cite{Staudt2014}. 
All experiments were done on a machine equipped with 256 GB RAM and a 2.7 GHz Intel Xeon CPU E5-2680 having 2 sockets with 8 cores each.
The machine runs 64 bit SUSE Linux and we compiled our code with g++-4.8.1 and OpenMP~3.1. For comparability with previous work, unless stated explicitly, running times refer to a sequential implementation.

The graphs used in the experiments are taken from the SNAP~\cite{snapnets}, KONECT~\cite{DBLP:conf/www/Kunegis13} and LASAGNE\footnote{\url{piluc.dsi.unifi.it/lasagne}} data sets. The \texttt{easyjet} graph in Table~\ref{table:accuracy} was taken from~\cite{DBLP:conf/wea/CrescenziDSV15}. All graphs are connected, undirected and unweighted.

\subsection{Accuracy}
\label{experiments:ilp}
The quality comparison between the \textsf{Greedy++} solution and the optimum is performed on several small real-world networks; the optimum is computed using the ILP formulation described in Section~\ref{sec:ilp}. 
The ILP model is implemented using the Java optimization modeling library and interface ILOG Concert Technology. The problems are solved with ILOG CPLEX 12.6\footnote{\url{www-01.ibm.com/software/commerce/optimization/cplex-optimizer/}}.
The results for $k=10$ are reported in Table~\ref{table:accuracy} in Appendix~\ref{app:results}. Among all networks, the empirical approximation ratio (ratio between the objective function of the optimum and that of the solution found by \textsf{Greedy++}) is always higher than 0.97. 
Similar results can be observed for $k=2$ and $k=20$, reported in Table~\ref{table:accuracy2} and Table~\ref{table:accuracy20} in Appendix~\ref{app:results}. For $k=10$, the geometric mean of the approximation ratios is 0.994, for $k=2$ it is 0.998 and for $k=20$ it is 0.995.
Notice that \textsf{Greedy++} never takes more than one second on the tested networks, whereas finding the optimum
with CPLEX takes hours for the larger instances of Table~\ref{table:accuracy}.

\subsection{Algorithmic speedup on \textsf{Greedy}} 
\label{sec:speedup}
\begin{figure}[t]
\centering
\includegraphics[width = 0.45\textwidth]{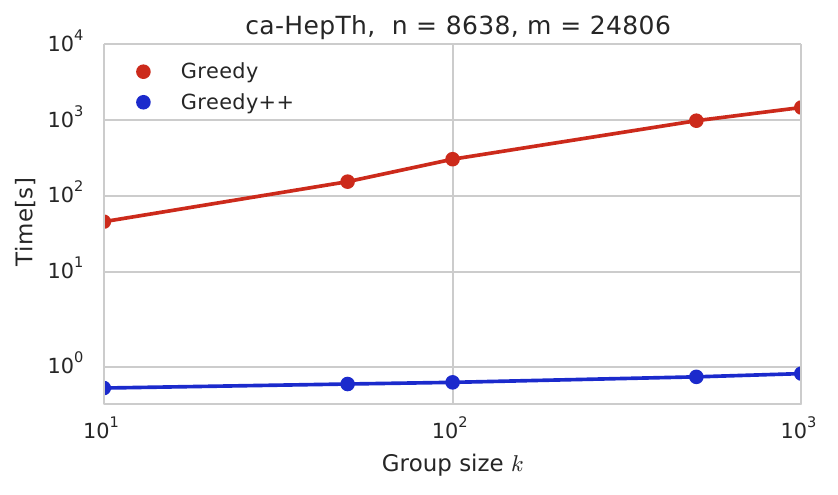}
\includegraphics[width = 0.45\textwidth]{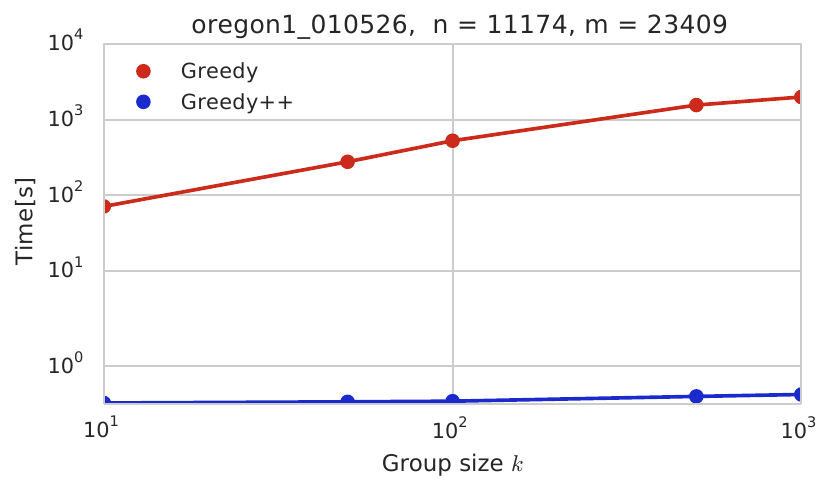}
\caption{Running times of \textsf{Greedy} and \textsf{Greedy++} for different group sizes (log-log scale). Top: running times for \texttt{ca-HepTh};
bottom: running times for \texttt{oregon\_1\_010526}.}
\vspace{-3ex}
\label{fig:compGreedy}
\end{figure}

Recall that the solution found by the two algorithms \textsf{Greedy++} and \textsf{Greedy} is the same, thus we only compare running times between the two. Due to the time and space complexity of \textsf{Greedy}, we compare the two approaches on two relatively small networks (\texttt{ca-HepTh}: 8638 nodes and 24806 edges and \texttt{oregon\_1\_010526}: 11174 nodes and 23409 edges).
Figure~\ref{fig:compGreedy} shows the running times of the two algorithms for different values of group size $k$ between 10 and 1000.
 For both graphs, \textsf{Greedy++} outperforms \textsf{Greedy} by orders of magnitude. For all tested group sizes, \textsf{Greedy++} finds the solution in less than one second, whereas for $k =1000$ \textsf{Greedy} requires 25 minutes on the \texttt{ca-HepTh} graph and 34 minutes on the \texttt{oregon\_1\_010526} graph. The speedups of \textsf{Greedy++} on \textsf{Greedy} ranges between 93 ($k=10$) and 1765 ($k=1000$) for \texttt{ca-HepTh} and between 581 ($k=10$) and 6125 ($k=1000$) for  \texttt{oregon\_1\_010526}.

\subsection{Comparison with \textsf{OSA}}
\label{sec:osa}
\begin{figure*}[tbh]
\begin{subfigure}[a]{\textwidth}
\begin{center}
\includegraphics[width = 0.45\textwidth]{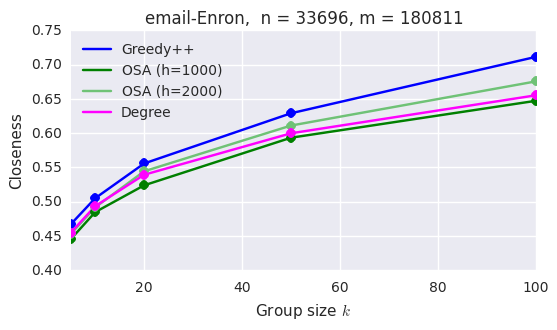}
\includegraphics[width = 0.45\textwidth]{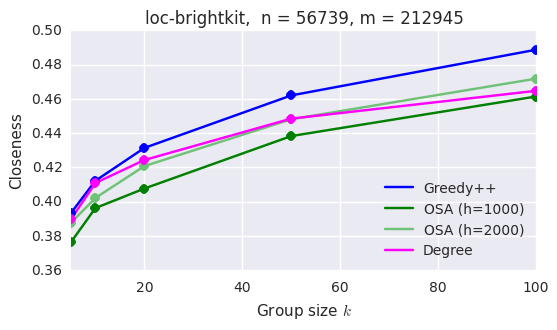}
\end{center}
\end{subfigure}

\begin{subfigure}[b]{\textwidth}
\begin{center}
\includegraphics[width = 0.45\textwidth]{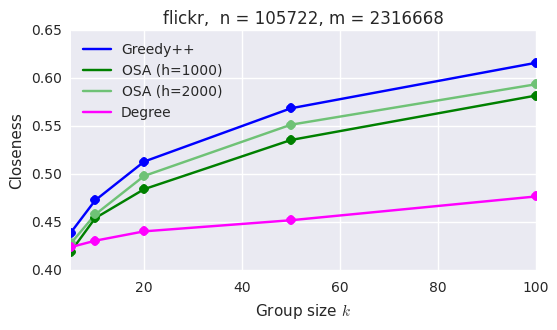}
\includegraphics[width = 0.45\textwidth]{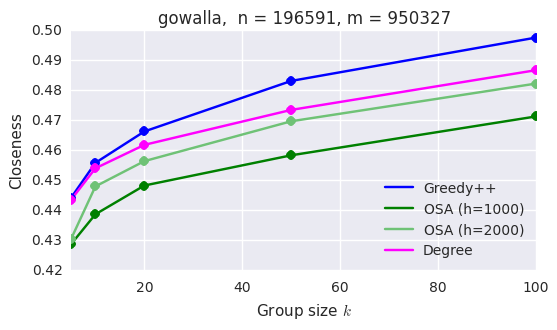}
\end{center}
\end{subfigure}
\caption{Closeness centrality of the solution found by the methods for different group sizes and different graphs. The plot shows the results of \textsf{Greedy++}, \textsf{OSA} with sample sizes of 1000 and 2000, and the group consisting of the $k$ nodes with highest degree.}
\label{fig:compOSA}
\vspace{-2ex}
\end{figure*}
Since \textsf{OSA} is a sampling-based algorithm, the number $h$ of samples influences its performance, both in terms of accuracy and running time. 
In~\cite{DBLP:conf/adc/ChenWW16}, the authors suggest $h=1000$ samples as a good tradeoff for group sizes up to 50. Since we are also testing the algorithms on groups with up to 100 nodes, we run \textsf{OSA} both with $h=1000$ and with a larger sample size of $h=2000$. 
We test \textsf{OSA} and \textsf{Greedy++} on all the networks of Table~\ref{table:networks} with $m < 10^7$ (11 networks). We did not run experiments on larger networks because of the high memory requirements of \textsf{OSA}. Since \textsf{OSA} is a sampling-based approach, we repeat each experiment 10 times and report the average running time and accuracy.
Figure~\ref{fig:compOSA} shows the group closeness of the solutions found by \textsf{OSA} and \textsf{Greedy++} on four of the tested graphs (\texttt{email-Enron}, \texttt{loc-brightkit}, \texttt{flickr}, and \texttt{gowalla}), for group sizes ranging between 5 and 100. As a baseline, we also report the closeness of the group composed of the $k$ nodes with maximum degree (\textsf{Degree}). 
%
The results show that \textsf{Greedy++} always finds a better solution, for all graphs and group sizes. Interestingly, for all the four graphs but \texttt{flickr}, the set of nodes with maximum degree has a higher closeness 
than the solution found by \textsf{OSA} with $h=1000$ samples. For the \texttt{gowalla} graph, \textsf{Degree} finds a better solution than \textsf{OSA} even with $h=2000$ samples.
Figure~\ref{fig:timesOSA} in Appendix~\ref{app:results} shows the running times of \textsf{Greedy++} and \textsf{OSA} on the four graphs, for group size $k=20$ (top) and $k=100$ (bottom). On all graphs but \texttt{flickr}, \textsf{Greedy++} is significantly faster than \textsf{OSA} (both with $h=1000$ and $h=2000$ samples). On the \texttt{flickr} graph, for group size $k=20$, \textsf{Greedy++} takes 85 seconds, whereas \textsf{OSA} with $h=2000$ takes 77 seconds. However, when the group size increases ($k=100$),  \textsf{Greedy++} becomes faster (102 seconds versus 182 seconds required by \textsf{OSA} with $h=2000$). 
Also, notice that the memory requirement of \textsf{Greedy++} is significantly lower than that of \textsf{OSA}. In fact, \textsf{Greedy++} only needs $\Theta(n)$ memory for its data structures, whereas \textsf{OSA} requires $\Theta(h n)$ to store the distances between the sampled nodes and the other nodes. This means that, using \textsf{OSA} with the number of samples suggested in~\cite{DBLP:conf/adc/ChenWW16}, it needs roughly one thousand times more memory than \textsf{Greedy++}, which might be problematic for large graphs.
On average (geometric mean) over the 11 tested networks, \textsf{Greedy++} is faster than \textsf{OSA} with $h=1000$ by a factor of 1.1 and than \textsf{OSA} with $h=2000$ by a factor of 1.7. Although our average running times are not very different from those of \textsf{OSA} with $h=1000$, our accuracy is better on all tested networks (the same is true also for \textsf{OSA} with $h = 2000$). Also, on 7 out of the 11 tested networks, \textsf{OSA} with $h=1000$ returns a result with a worse accuracy than choosing the $k$ nodes with maximum degree, suggesting that \textsf{OSA} should be run using a larger number of samples. With $h=2000$, the solution of \textsf{OSA} is worse than picking the $k$ nodes with maximum degree on 4 out of 11 networks (the solution returned by \textsf{Greedy++} is better on all tested networks).

\subsection{Running time evaluation}
\label{sub:running-time}
To test the scalability of \textsf{Greedy++}, we now run it on all networks from Table~\ref{table:networks} (for the comparison with \textsf{OSA}, only the first 11 networks could be used). The networks belong to different domains, including friendship, collaboration, communication and internet topology graphs.
To further speed up the running time of \textsf{Greedy++}, we also implement a parallel version of it. The first element of $|S|$ is computed using the parallel top-$k$ closeness implementation described in~\cite{DBLP:journals/corr/BergaminiBCMM17}. Then, in each iteration of \textsf{Greedy++}, Line~\ref{line:visitNodes} of Algorithm~\ref{alg:mem-greedy} in Appendix~\ref{app:pseudocodes} is executed in parallel, i.e. each thread runs a pruned SSSP from the nodes assigned to it.
Table~\ref{table:networks} reports the running times of \textsf{Greedy++} for $k=10$, for both the sequential and the parallel implementation (using 16 threads). On all networks with less than $10^5$ nodes, our parallel implementation takes less than 1 second. On all remaining graphs, it always takes less than 1 hour, apart from the \texttt{com-orkut} graph ($> 3$M nodes and $> 100$M edges), where it takes a bit more than one and a half hours.
The parallel speedup varies significantly among the tested networks, ranging from 5.4 (\texttt{com-youtube}) to 13.8 (\texttt{flickr}). These values should be appreciated in the context of complex networks, for which it is
often difficult to obtain even higher speedups (see for example~\cite{DBLP:conf/hipc/McCollGB13} and~\cite{DBLP:journals/tpds/StaudtM16}).
Low speedup values are in our case also due to the fact that, in some networks, the work done by the pruned SSSPs is extremely imbalanced (some nodes can be pruned early, whereas others need almost a full SSSP).
Load balancing mechanisms beyond what OpenMP offers are outside the scope of this paper, as they would require very fine-grained and inexpensive context switches between threads.
Also, as expected, the parallel speedup decreases as $k$ increases. Indeed, whereas the geometric mean of the speedups is $9.1$ for $k = 10$, it is $8.7$ for $k = 20$ and $5.6$ for $k = 100$. This results from the fact that, for higher values of $k$, more and more pruned SSSPs can be skipped because of supermodularity. Since less work is done in each iteration, the overhead due to parallelism and imbalance becomes more significant.
The fact that less and less work is done in each iteration as $k$ increases is also confirmed by the fact that the running times do not increase linearly as $k$ increases. For $k=20$, the running times are only about $10\%$ higher (on average) that they are for $k = 10$ and, for $k = 100$, they are about $50\%$ higher than for $k=10$ (running times for $k = 20$ and $k= 100$ can be found in Table~\ref{table:times20-100} in Appendix~\ref{app:results}).

\begin{table*}[tbh]
  \centering
  \caption{Networks used in the experiments and performance of \textsf{Greedy++} for $k = 10$. The fourth and fifth columns report the sequential and parallel running times with 16 threads, respectively. The last column reports the speedup of the parallel implementation on the sequential one.}

\begin{small}
  \begin{tabular}{  | l | r | r | r | r | r |}
\hline
Graph & Nodes & Edges & Time seq. [s] & Time par. [s] & Speedup  \\
\hline
\texttt{ca-HepPh}  &  11204  &  117649 & 7.70 & 0.58 & 13.4 \\
\texttt{email-Enron}  &  33696  &  180811 &1.94 & 0.20 & 9.9 \\
\texttt{CA-AstroPh}  &  17903  &  197031 & 3.78 & 0.32 & 12.0 \\
\texttt{loc-brightkite}  &  56739  &  212945 &5.74 & 0.55 & 10.5 \\
\texttt{com-lj}  &  303526  &  427701 &127.35 & 17.00 & 7.5 \\
\texttt{com-amazon}  &  334863  &  925872 & 808.70 & 88.37 & 9.2 \\ 
\texttt{gowalla}  &  196591  &  950327 & 60.14 & 8.74 & 6.9 \\
\texttt{com-dblp}  &  317080  &  1049866 & 232.51 & 30.99 & 7.5 \\
\texttt{flickr}  &  105722  &  2316668 & 314.11 & 22.76 & 13.8 \\
\texttt{com-youtube}  &  1134890  &  2987624 & 1323.31 & 245.50 & 5.4 \\
\texttt{youtube-u-growth}  &  3216075  &  9369874 & 22298.52 & 2196.42 & 10.2 \\
\texttt{as-skitter}  &  1694616  &  11094209 & 12014.09 & 1611.09 & 7.5 \\
\texttt{soc-pokec-relationships}  &  1632803  &  22301964 & 11912.29 & 1104.82 & 10.8 \\
\texttt{com-orkut}  &  3072441  &  117185083 & 60252.10 & 5792.81 & 10.4 \\ 
\hline
\end{tabular}
\vspace{-2ex}
\label{table:networks}
\end{small}
\end{table*}

\subsection{\textsf{Greedy++} using bit vectors}
\label{sec:exp-bitwise}
We now test the performance of the version of \textsf{Greedy++} using bit vectors described in Section~\ref{sec:bitwise}. Our implementation of bit vectors is based on the C++ \texttt{std::bitset} and we test both versions sequentially. Table~\ref{table:bitwise} in Appendix~\ref{app:results} shows the ratio between the running times of \textsf{Greedy++} using pruned SSSPs and \textsf{Greedy++} using bit vectors, for $k = 10$, $k =100$ and $k = 1000$ (we call the version using bit vectors \textsf{bitGreedy++}). The ratio is never smaller than $0.9$ and \textsf{bitGreedy++} is up to a factor $4$ faster than \textsf{Greedy++}. The geometric means of the ratios are $1.1$ for $k = 10$, $1.6$ for $k = 100$ and $2.8$ for $k = 1000$. On the other hand, the memory required by \textsf{bitGreedy++} is usually much higher. On \texttt{com-amazon}, \textsf{Greedy++} requires only about 312 MB, whereas \textsf{bitGreedy++} needs 226 GB. For this reason, we were not able to test \textsf{Greedy++} on the 5 largest networks of Table~\ref{table:networks}.
To summarize, \textsf{bitGreedy++} is mostly faster than \textsf{Greedy++}, and the improvement is more apparent for larger values of $k$. Thus, using \textsf{bitGreedy++} is recommended if enough memory is available and $k$ is relatively large (e.g. $k\geq 100$).

\subsection{Group closeness versus top-$k$ closeness}
\label{sec:group-vs-top}
A natural question is how many elements of the group of nodes with highest closeness have high closeness or high degree individually. We investigate this on the networks of Table~\ref{table:networks}. In particular, for a given group size $k$, we compute the overlap (\ie the size of the intersection) between the group returned by \textsf{Greedy++} and the set of the top-$k$ nodes with highest closeness (computed using the algorithm described in~\cite{DBLP:conf/alenex/BergaminiBCMM16}, which is available in NetworKit) and highest degree. The percentage overlap is then the overlap divided by $k$ and multiplied by 100. 
Figure~\ref{fig:compTopK} in Appendix~\ref{app:results} shows the results. The plot on the bottom right corner shows the average over all networks of Table~\ref{table:networks}, whereas the other three plots refer to the \texttt{com-youtube} graph, to \texttt{soc-pokec-relationships} and to \texttt{com-orkut}, respectively. As it appears from the plots, the overlap changes significantly among the graphs. For the \texttt{com-youtube} graph, the percentage overlap decreases as the group size increases, and the overlap with \textsf{Degree} is always larger than the one with \textsf{Top-$k$}. Partially similar are the results for \texttt{soc-pokec-relationships}, although there is more fluctuation in the overlap of \textsf{Degree} and the initial overlap of  \textsf{Top-$k$} is higher than it is for \texttt{com-youtube} ($\approx 80\%$ vs. $\approx 60\%$). On the other hand, the results for \texttt{com-orkut} are quite different: The overlap with \textsf{Degree} increases with the group size, and is lower than the one with \textsf{Top-$k$}. 
On average, the overlap with both \textsf{Degree} and \textsf{Top-$k$} tends to decrease as the group size increases (as expected), with \textsf{Degree} having a higher overlap than \textsf{Top-$k$} (except for $k = 5$). Also, on average the overlap ranges between $30\%$ and $60\%$. This clearly indicates that there is a dependence between the group with maximum closeness and the degrees of nodes and their centralities. However, the strength of this dependence varies significantly among the tested networks and suggests that picking the $k$ nodes with highest closeness or highest degree is not always a good heuristic for finding the group with maximum closeness.

\section{Conclusions}
In this paper we have studied the problem of finding the group with maximum closeness in large complex networks. 
Our algorithm scales to networks with tens or hundreds of millions of edges \emph{and} delivers 
an excellent \emph{empirical} approximation ratio at the same time (never lower than $0.97$ on the networks
for which we could compute the optimal solution in reasonable time).
Pruning the SSSP searches and exploiting the supermodularity of farness allows us to reduce the amount of work done by the greedy algorithm proposed in~\cite{DBLP:conf/adc/ChenWW16} by orders of magnitude.
%
Also, using our approach, we have been able to study the relation between a group with high closeness and nodes that have individually high closeness or degree in large complex networks.

Future work includes an extension of our approach to disk-resident graphs, for which a comparison with the heuristic proposed in~\cite{DBLP:conf/www/ZhaoLTG14} would be interesting. 
%
It would also be interesting to study how an extension of our greedy algorithm would perform on the $p$-median problem with node weights.

\paragraph*{Acknowledgments}
We are very grateful to Eugenio Angriman and Alexander van der Grinten (both HU Berlin) for providing
the counterexample we presented in Section~\ref{sec:greedy} regarding the submodularity of group closeness  
and for other helpful discussions on the topic.

\FloatBarrier
%
\bibliographystyle{abbrv}
\balance 
\bibliography{references}  
%
%


\newpage
\appendix

\section{Additional pseudocodes}
\label{app:pseudocodes}

\begin{algorithm2e}[h]
\LinesNumbered
\SetKwFunction{BFScut}{BFScut}
\SetKwFunction{Kth}{Kth}
\SetKwFunction{TopK}{TopK}
\SetKwFunction{APSP}{APSP}
\SetKwData{farn}{Farn}
\SetKwData{Q}{Q}
\SetKwData{L}{L}
\SetKwData{Top}{Top}
 \SetKwInOut{Input}{Input}
 \SetKwInOut{Output}{Output}
\Input{A graph $G=(V,E)$, a number $k$}
\Output{A set $S \subseteq V$ of size $k$}
$d \leftarrow \APSP(G)$\; 
$M \leftarrow \APSP(G)$\; \label{line:distances}
$Score \leftarrow \{c(u) | u \in V \}$\;
$s \leftarrow {\arg \max}_{u \in V \setminus S} Score[u]$\;
$S \leftarrow  \{s\}$\;
\While{$|S| < k $}{
\ForEach{$u \in  V \setminus S$}{
	\ForEach{$w \in  V$}
	{ \label{line:for1}
		\If{$d[u, w] > d[s, w]$}{
			$M[u, w] \leftarrow d[s, w]$\; \label{line:updateM}
		}
	}
	\tcc{$Score[u] $ is set to $c(S \cup \{u\})$}
	$Score[u] \leftarrow (n-|S|-1)/\sum_{w \in V\setminus S} M[u, w]$\; \label{line:updateScore}
	$s \leftarrow {\arg \max}_{w \in V \setminus S} Score[w]$\;
	$S \leftarrow S \cup \{s\}$\;
}
}
\Return $S$\;
\caption{Greedy algorithm for GCM~\cite{DBLP:conf/adc/ChenWW16}.}
\label{alg:greedy}
\end{algorithm2e}

\begin{algorithm2e}[h!]
\LinesNumbered
\SetKwFunction{BFScut}{BFScut}
\SetKwFunction{SSSP}{SSSP}
\SetKwFunction{TopKCloseness}{TopKCloseness}
\SetKwData{farn}{Farn}
\SetKwData{Q}{Q}
\SetKwData{L}{L}
\SetKwData{Top}{Top}
 \SetKwInOut{Input}{Input}
 \SetKwInOut{Output}{Output}
\Input{A graph $G=(V,E)$, a number $k$}
\Output{A set $S \subseteq V$ of size $k$}
$s_0 \leftarrow \TopKCloseness(1)$\; \label{line:topcloseness}
$S \leftarrow \{ s_0 \}$\;
$\SSSP(s_0)$\; \label{line:sssp0}
$d_S[u] \leftarrow d(s_0, u)\ \  \forall u \in V$\;
\While{$|S| < k $}{
\label{line:while}
\ForEach{$u \in  V \setminus S$}{
\label{line:visitNodes}
	$\SSSP(u)$\; \label{line:sssp}
	\tcc{$Score[u] $ is set to $c(S \cup \{u\})$}
	$t \leftarrow \sum_{w \in V\setminus S} \min\{d(u, w),\  d_S[w]\}$\; \label{line:sumScore}
	$Score[u] \leftarrow (n-|S|-1)/t$\;  \label{line:actScore}
}
$s \leftarrow {\arg \max}_{w \in V \setminus S} Score[w]$\; \label{line:nodeMax}
$S \leftarrow S \cup \{s\}$\;
$\SSSP(s)$\; \label{line:sssp2}
\ForEach{$u \in V$}{
$d_S[u] \leftarrow \min\{d_S[u], \ d(s, u) \}$\; \label{line:compdS}
}
}
\Return $S$\;
\caption{Memory-efficient greedy algorithm, here expressed in terms of closeness.
  Our implementation is based on the farness formulation. Recall from Section~\ref{sec:submodularity}
  that exploiting supermodularity does not change the solution unless two nodes have the same marginal gain.}
\label{alg:mem-greedy}
\end{algorithm2e}

\begin{algorithm2e}[h]
\LinesNumbered
\SetKwFunction{BFScut}{BFScut}
\SetKwFunction{Kth}{Kth}
\SetKwFunction{TopK}{TopK}
\SetKwData{farn}{Farn}
\SetKwData{Q}{Q}
\SetKwData{L}{L}
\SetKwData{Top}{Top}
 \SetKwInOut{Input}{Input}
 \SetKwInOut{Output}{Output}
\Input{A graph $G=(V,E)$, a number $k$}
\Output{Top $k$ nodes with highest closeness}
$c(v) \leftarrow 0 \ \ \forall v \in V$\;
$x_k \leftarrow 0$\;
\For{$v \in V$ in decreasing order of degree}{
$c(v) \leftarrow \BFScut(v, x_k)$\; \label{line:bfscut}
\If{$c(v) \neq 0$}{
	$x_k \leftarrow \Kth(c)$\; \label{line:kth}
}
}
\Return $\TopK(c)$\; \label{line:topk}
\caption{Top-$k$ closeness centrality~\cite{DBLP:journals/corr/BergaminiBCMM17}.}
\label{alg:topk}
\end{algorithm2e}
\newpage
\FloatBarrier

\section{Additional experimental results}
\label{app:results}
\FloatBarrier
\begin{table*}[tb]
  \centering
  \caption{Comparison with optimum on small real-world networks, for $k = 10$. The fourth and fifth columns show the objective function of Eq.~(\ref{eq:lp2}) for the optimum and \textsf{Greedy++}, respectively.}
  \begin{tabular}{  | l | r | r | r | r | r | r |}
\hline
Graph & Nodes & Edges & Category & Optimum & \textsf{Greedy++} & Approx. ratio\\
\hline
\texttt{karate} &  35  &  78 & friendship & 25 & 25 & 1.0\\ 
\texttt{contiguous-usa} &  49  &  107 & transport. & 40 & 41 & 0.976 \\ 
\texttt{easyjet} &  136  &  755 & transport. & 126 & 126 & 1.0 \\ 
\texttt{jazz} &  198  &  2742 & collaboration & 191 & 192 & 0.995\\ 
\texttt{coli1-1Inter} &  328  &  456 & metabolic & 475 & 482 & 0.985\\ 
\texttt{pro-pro} &  1458  &  1993 & metabolic  & 4213 & 4217& 0.999  \\ 
\texttt{hamster-friend} &  1788  &  12476 & social  & 2871 & 2871 & 1.0  \\ 
\texttt{dnc-temporal} &  1833  &  4366 & communicat. & 2398 & 2407  &  0.996 \\ 
\texttt{caenorhab-eleg} &  4428  &  9659 & metabolic  & 10003 & 10075 & 0.993  \\ 

\hline
\end{tabular}
\label{table:accuracy}
\end{table*}

\begin{table*}[bth]
  \centering
  \caption{Comparison with optimum on small real-world networks, for $k = 2$. The fourth and fifth columns show the objective function of Eq.~(\ref{eq:lp2}) for the optimum and \textsf{Greedy++}, respectively.}
  \begin{tabular}{  | l | r | r | r | r | r | r |}
\hline
Graph & Nodes & Edges & Category & Optimum & \textsf{Greedy++} & Approx. ratio\\
\hline
\texttt{karate} &  35  &  78 & friendship & 37 & 37 & 1.0 \\ 
\texttt{contiguous-usa} &  49  &  107 & transport. & 99 & 99 & 1.0  \\ 
\texttt{easyjet} &  136  &  755 & transport. & 143 & 143 & 1.0 \\ 
\texttt{jazz} &  198  &  2742 & collaboration & 259 & 261  & 0.992\\ 
\texttt{coli1-1Inter} &  328  &  456 & metabolic & 780 &  780 & 1.0\\ 
\texttt{pro-pro} &  1458  &  1993 & metabolic  & 5573 & 5573 & 1.0  \\ 
\texttt{hamster-friend} &  1788  &  12476 & social  & 3596 & 3596 & 1.0  \\ 
\texttt{dnc-temporal} &  1833  &  4366 & communicat. & 3236 & 3236 &  1.0 \\ 
\texttt{caenorhab-eleg} &  4428  &  9659 & metabolic  & 12535 & 12631 & 0.992  \\ 

\hline
\end{tabular}
\label{table:accuracy2}
\end{table*}

\begin{table*}[bth]
  \centering
  \caption{Comparison with optimum on small real-world networks, for $k = 20$. The fourth and fifth columns show the objective function of Eq.~(\ref{eq:lp2}) for the optimum and \textsf{Greedy++}, respectively. The results for \texttt{caenorhab-eleg} are not included, because the CPLEX solver did not find the optimum within 13 hours.}
  \begin{tabular}{  | l | r | r | r | r | r | r |}
\hline
Graph & Nodes & Edges & Category & Optimum & \textsf{Greedy++} & Approx. ratio\\
\hline
\texttt{karate} &  35  &  78 & friendship & 15 & 15 & 1.0  \\ 
\texttt{contiguous-usa} &  49  &  107 & transport. & 29 & 29 & 1.0  \\ 
\texttt{easyjet} &  136  &  755 & transport. & 116 & 116 & 1.0 \\ 
\texttt{jazz} &  198  &  2742 & collaboration & 178 & 178  & 1.0\\ 
\texttt{coli1-1Inter} &  328  &  456 & metabolic & 367 & 373  & 0.984\\ 
\texttt{pro-pro} &  1458  &  1993 & metabolic  & 3488 & 3518 &  0.991 \\ 
\texttt{hamster-friend} &  1788  &  12476 & social  & 2556 & 2573 &  0.993 \\ 
\texttt{dnc-temporal} &  1833  &  4366 & communicat. & 2066 & 2082 &  0.992 \\ 

\hline
\end{tabular}
\label{table:accuracy20}
\end{table*}

\begin{figure}[th]
\centering
\includegraphics[width = 0.45\textwidth]{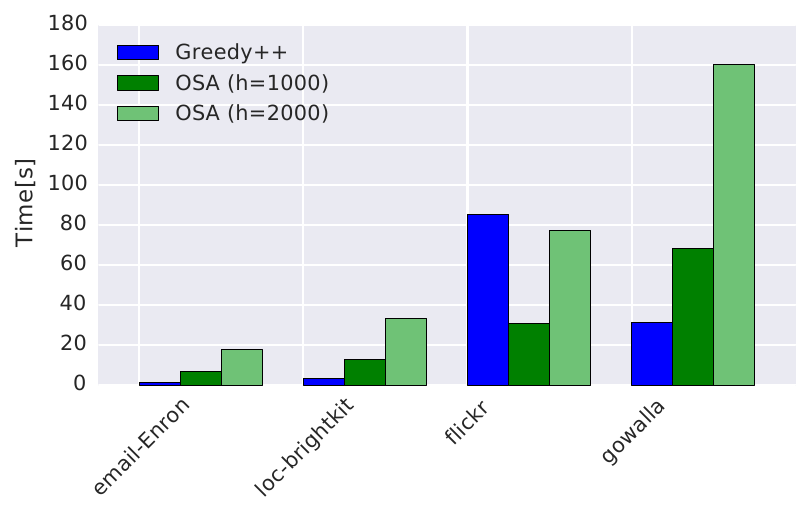}
\includegraphics[width = 0.45\textwidth]{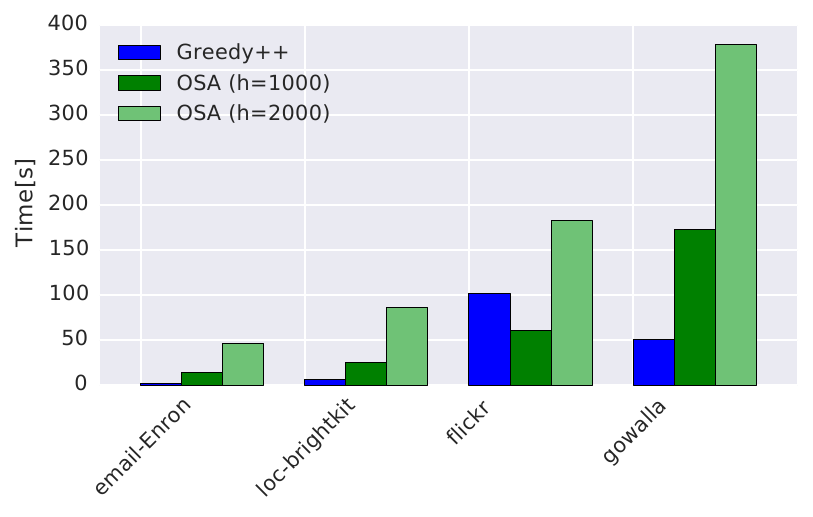}
\caption{Running times of \textsf{Greedy++} and \textsf{OSA} with sample sizes of 1000 and 2000 for $k=20$ (top) and $k=100$ (bottom).}
\label{fig:timesOSA}
\end{figure}

\begin{table*}[tbh]
  \centering
  \caption{Performance of \textsf{Greedy++} for $k = 20$ and $k = 100$, using 16 threads.}
  \begin{tabular}{  | l | r | r | r | r | }
\hline
Graph & Nodes & Edges & Time $k=20$ [s] & Time $k=100$ [s]  \\
\hline
\texttt{ca-HepPh}  &  11204  &  117649 & 0.61 & 0.7 \\
\texttt{email-Enron}  &  33696  &  180811 &0.26 & 0.6 \\
\texttt{CA-AstroPh}  &  17903  &  197031 & 0.34 & 0.5 \\
\texttt{loc-brightkite}  &  56739  &  212945 &0.67 & 1.2 \\
\texttt{com-lj}  &  303526  &  427701 &18.16 & 24.2 \\
\texttt{com-amazon}  &  334863  &  925872 & 94.56 & 116.2 \\
\texttt{gowalla}  &  196591  &  950327 & 9.09 & 11.2 \\
\texttt{com-dblp}  &  317080  &  1049866 & 34.26 & 49.5 \\
\texttt{flickr}  &  105722  &  2316668 & 23.04 & 24.6 \\
\texttt{com-youtube}  &  1134890  &  2987624 & 263.17 & 473.9 \\
\texttt{youtube-u-growth}  &  3216075  &  9369874 & 2412.60 & 2901.9 \\
\texttt{as-skitter}  &  1694616  &  11094209 & 1620.43 & 2024.6 \\
\texttt{soc-pokec-relationships}  &  1632803  &  22301964 & 1179.33 & 1288.1 \\
\texttt{com-orkut}  &  3072441  &  117185083 & 6233.67 & 8387.0 \\
\hline
\end{tabular}
\label{table:times20-100}
\end{table*}

\begin{figure*}[h]
\begin{subfigure}[a]{\textwidth}
\begin{center}
\includegraphics[width = 0.45\textwidth]{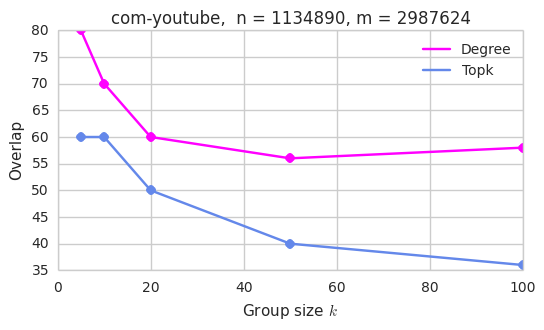}
\includegraphics[width = 0.45\textwidth]{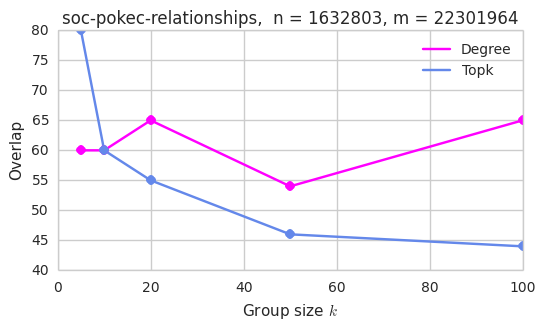}
\end{center}
\end{subfigure}

\begin{subfigure}[b]{\textwidth}
\begin{center}
\includegraphics[width = 0.45\textwidth]{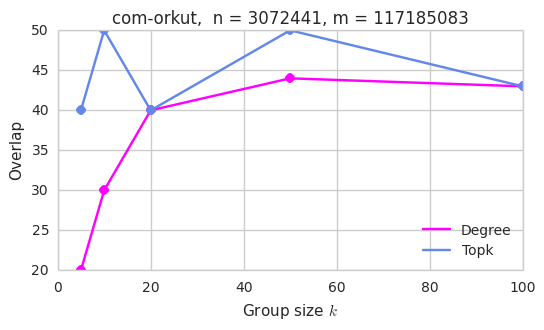}
\includegraphics[width = 0.45\textwidth]{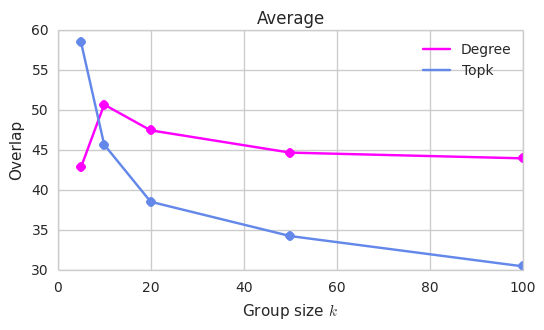}
\end{center}
\end{subfigure}
\caption{Percentage overlap between the group found by \textsf{Greedy++} and the $k$ nodes with highest closeness (\textsf{Top-$k$}) and between the group found by \textsf{Greedy++} and the $k$ nodes with highest degree (\textsf{Degree}).}
\label{fig:compTopK}
\end{figure*}

\begin{table*}[tbh]
  \centering
  \caption{Performance of the new algorithm for group closeness using pruned SSSPs (\textsf{Greedy++}) and using bit vectors (\textsf{bitGreedy++}). The first three columns represent the speedup of \textsf{bitGreedy++} on \textsf{Greedy++} (i.e. the ratio between their running times). The last two columns report the memory requirements.}
  \begin{tabular}{  | l | p{1.7 cm} | p{1.7 cm} | p{1.7 cm}   | l | l |}
    \cline{2-4}
\multicolumn{1}{c|}{} & \multicolumn{3}{c|}{Speedup of \textsf{bitGreedy++} on \textsf{Greedy++}} \\
\hline
Graph &  $k = 10$ & $k = 100$ & $k = 1000$  & Mem. \textsf{Greedy++} & Mem. \textsf{bitGreedy++} \\
\hline
\texttt{ca-HepPh}  &  0.95  &  1.59  &  3.90  & $\approx 136$ MB &  $\approx 210$ MB  \\
\texttt{email-Enron}  &  1.16  &  1.74  &  4.09  & $\approx 151$ MB  & $\approx 603$ MB \\
\texttt{CA-AstroPh}  &  0.90  &  1.48  &  3.08 & $\approx 164$ MB &  $\approx 367$ MB  \\
\texttt{loc-brightkite}  &  0.97  &  1.39  &  3.00 & $\approx 273$ MB &  $\approx 1$ GB  \\
\texttt{com-lj}  &  0.96  &  1.24  &  2.10 & $\approx 318$ MB &  $\approx 78$ GB  \\
\texttt{com-amazon}  &  0.97  &  1.46  &  2.12 & $\approx 312$ MB  &  $\approx 226$ GB  \\
\texttt{gowalla-edges}  &  1.35  &  1.52  &  2.45 & $\approx 279$ MB &  $\approx 17$ GB  \\
\texttt{com-dblp}  &  1.13  &  1.70  &  2.57 & $\approx 310$ MB &  $\approx 94$ GB  \\
\texttt{flickrEdges}  &  1.60  &  2.04  &  2.73 & $\approx 339$ MB &  $\approx 5$ GB  \\
\hline
\end{tabular}
\label{table:bitwise}
\end{table*}

\end{document}